\documentclass[a4paper,11pt]{article}

\usepackage{authblk}
\usepackage{graphicx}
\usepackage{float}
\usepackage[version=3]{mhchem} 
\usepackage[utf8]{inputenc}
\usepackage[dvipsnames,table,xcdraw]{xcolor}
\usepackage{siunitx}
\mathchardef\mhyphen="2D
\usepackage{upgreek}
\usepackage{amsmath}
\usepackage{eucal}
\usepackage{soul}
\usepackage{bm}

\usepackage{hyperref}
\hypersetup{
	colorlinks,
	linkcolor={blue!90!black},
	citecolor={blue!90!black},
	urlcolor=	{blue!90!black}
}
\let\oldeqref\eqref
\renewcommand*{\eqref}[1]{%
	\hyperref[#1]{\oldeqref{#1}}%
}

\newcommand*{\kp}{\bm{k}{\cdot}\bm{p}}

\newcommand{\subfigref}[2]{Fig.~\hyperref[fig:#1]{\ref*{fig:#1}#2}}
\newcommand{\Supplsubfigref}[2]{Fig.~\hyperref[fig:#1]{\ref*{fig:#1}#2}}
\newcommand{\subfigrefL}[2]{Figure~\hyperref[fig:#1]{\ref*{fig:#1}#2}}
\newcommand{\subfigsref}[3]{Figs.~\hyperref[fig:#1]{\ref*{fig:#1}#2}-\hyperref[fig:#1]{\ref*{fig:#1}#3}}
\newcommand{\Supplsubfigsref}[3]{Figs.~S\hyperref[fig:#1]{\ref*{fig:#1}#2}-S\hyperref[fig:#1]{\ref*{fig:#1}#3}}
\newcommand{\subfigsrefL}[3]{Figures~\hyperref[fig:#1]{\ref*{fig:#1}#2}-\hyperref[fig:#1]{\ref*{fig:#1}#3}}

\author[1,2]{Yury Berdnikov\thanks{yurber@dtu.dk}}
\author[2,3]{Pawe\l{} Holewa}
\author[1,4]{Shima Kadkhodazadeh}
\author[3]{Jan Miko\l{}aj \'{S}migiel}
\author[2]{Adrianna Fr{a}ckowiak}
\author[4]{Aurimas Sakanas}
\author[1,2]{Kresten Yvind}
\author[3]{Marcin Syperek}

\author[1,2]{Elizaveta Semenova}

\affil[1]{NanoPhoton - Center for Nanophotonics, Technical University of Denmark, Kongens Lyngby 2800, Denmark}
\affil[2]{DTU Electro, Technical University of Denmark, Kongens Lyngby 2800, Denmark}
\affil[3] {Laboratory for Optical Spectroscopy of Nanostructures, Wroc\l{}aw University of Science and Technology, 50-370 Wroc\l{}aw, Poland}
\affil[4]{DTU Nanolab - National Centre for Nano Fabrication and Characterization, Technical University of Denmark, Kongens Lyngby 2800, Denmark}

\date{}

\title{Fine-tunable near-critical Stranski-Krastanov growth of InAs/InP quantum dots}

\begin{document}
\maketitle

\begin{abstract}
  Emerging applications of self-assembled semiconductor quantum dot (QD)-based nonclassical light sources emitting in the telecom C-band ($\SIrange{1530}{1565}{\nano\meter}$) present challenges in terms of controlled synthesis of their low-density ensembles, critical for device processing with an isolated QD. This work shows how to control the surface density and size of InAs/InP quantum dots over a wide range by tailoring the conditions of Stranski-Krastanow growth. We demonstrate that in the near-critical growth regime, the density of quantum dots can be tuned between $10^7$ and $10^{10}$~$\mathrm{cm}^{-2}$. Furthermore, employing both experimental and modelling approaches, we show that the size (and therefore the emission wavelength) of InAs nanoislands on InP can be controlled independently from their surface density. Finally, we demonstrate that our growth method gives low-density ensembles resulting in well-isolated QD-originated emission lines in the telecom C-band.
\end{abstract}

\section{Introduction}

The size-dependent electronic structure and carrier dynamics of self-assembled semiconductor III-V quantum dots (QDs) make them attractive as active media of classical optoelectronic and photonic devices (such as lasers, optical amplifiers, memories and other) \cite{bhattacharya2015, garcia2021, yu2017, cotta2020}. At the same time, the rapidly developing applications of QDs as nonclassical quantum light sources set stringent conditions on their properties, and thus, presents new challenges in terms of their synthesis\cite{sun2018single, senellart2017,  orieux2017}.
Stress-driven, so-called Stranski-Krastanow (SK), method to grow semiconductor QDs is widely used for a broad range of device applications \cite{garcia2021,shchukin2004epitaxy, bhattacharya2015, tartakovskii2012}. This method has been studied theoretically and experimentally for several semiconductor materials including InAs/GaAs, InAs/InP, Si/Ge and others\cite{shchukin2004epitaxy, bhattacharya2015, tartakovskii2012}. However, the emerging device applications such as single QD-based nonclassical photon emitters or entangled-photon pairs sources, present new demands in terms of density and morphology of QD ensembles, which are not always compatible with the previous approaches \cite{garcia2021,orieux2017, Holewa2020PRB} and require substantial modifications to the growth conditions. In particular, quantum devices accessing individual QDs require highly tailored growth of low surface density QD ensembles, which has so far remained challenging using the SK method.

Theoretical models describe the SK formation of QDs as the phase transition between the strained 2D wetting layer (WL) and the partially relaxed 3D nanostructures \cite{venables1983, ratsch1996, joyce2004, evans2006, barabasi1999, li2010, dubrovskii2014}. 
According to the generally accepted theory, formation of the 3D islands becomes thermodynamically preferable to the 2D WL growth when the nominal thickness of the deposited layer (denoted as $h$) exceeds the equilibrium value (denoted as $h_{eq}$).
The difference between $h$ and $h_{eq}$ characterizes the supersaturation (or superstress) of the WL, which determines the height of the nucleation barrier, and thus, the rate of QDs nucleation \cite{li2010, dubrovskii2014, osipov2002stress}. Based on this model, the pace of transition from the WL to the QDs increases with the WL thickness (and strain accumulation) until the WL consumption rate is large enough to balance the material influx from the precursors of the growth species. This corresponds to the so-called "critical" thickness of the WL denoted as $h_c$ which limits the real achievable thickness of the WL during the QD nucleation. We note that the term "critical" here should not be confused with the critical layer thickness for the formation of dislocations.

In the case of a highly superstressed system, when $h$ reaches the critical value $h_c$, the phase transition has a “supercritical” character. This implies the fast nucleation of a high number density of QDs (typically $10^{10} - 10^{11}$  $cm^{-2} $), leading to a rapid decline in superstress and preventing further QDs nucleation \cite{osipov2002stress,dubrovskii2014}.

In contrast, when the thickness of the WL is below the critical value but exceeds the equilibrium value ($h_{eq} < h < h_c$), the phase transition has a subcritical character. In this case, nucleation of subcritical QDs occurs at lower supersaturation compared to the supercritical case and is thus much slower. As a result, the surface density of subcritical QDs is one or two orders of magnitude lower than in the supercritical case. However, continuous nucleation leads to a large difference in growth duration from QD to QD, and thus, results in a large dispersion in the size of the QD ensemble \cite{dubrovskii2014}.

Subcritical and supercritical nucleation have previously been studied theoretically and explored experimentally, mainly for the In(Ga,As)/GaAs material system \cite{song2006,seravalli2011single,li2013inas}. However, the intermediate nucleation at $h \approx h_c$, which we refer to as “near-critical”, remains less scrutinized. The systematic investigation of this regime for InAs QDs formation on GaAs substrates is difficult due to a rather small critical thickness $h_c \approx$ 1.7 ML, which is close to $h_{eq}$ \cite{song2006}.  In contrast, InAs/InP system has a larger difference between $h_c$ of around 4~ML and $h_{eq}$ of around 3 MLs according to recent experimental observations \cite{imec2019}.  

Furthermore, in contrast to the In(Ga,As) case, the thickness of the In(As,P) layer on top of InP can be tuned not only by the layer deposition but also by employing the interchange of P and As atoms at the interface during the annealing under the flux of arsenic precursor. 

In this work we demonstrate how to obtain a high degree of control over the surface density and the size distribution of SK InAs/InP QDs in a wide range by tailoring the growth parameters. InAs QDs on InP(100) surface densities of $ 10^{10} - 10^{11}$~$\mathrm{cm}^{-2}$ are readily accessible within the supercritical growth, while in the near-critical regime we show how the low density of QDs can be tuned in a very wide range between $10^6$ and $10^{10}$~$\mathrm{cm}^{-2}$. Furthermore, we explain how the size of InAs nanoislands on InP can be controlled independently from the surface density. Moreover, we measure the photoluminescence response from the low-density QDs ensemble, revealing sharp and well-isolated emission lines within the C-band originating from recombination of excitonic complexes confined to individual QDs.

\section{Methods}

\subsection{Epitaxial growth}
All the samples were grown in a low-pressure metal-organic vapor phase epitaxy (MOVPE) Turbodisc\textregistered \: reactor on InP wafers with (001) orientation. 
H$_2$  was used as a carrier gas, trimethylindium (TMIn) as the In source, phosphine (PH$_3$), tertiarybutylphosphine (TBP), and arsine (AsH$_3$) were used as precursors of the V\textsuperscript{th} group.
After thermal deoxidation of the InP(001) substrate at $\SI{650}{\degreeCelsius}$ in PH$_3$ ambient, the temperature was decreased to $\SI{610}{\degreeCelsius}$ and a $\SI{0.5}{\micro\meter}$-thick InP buffer layer was deposited. Then the temperature was reduced to $\SI{485}{\degreeCelsius}$ for QD growth under PH$_3$ with following annealing in AsH$_3$ ambient prior to InAs deposition. After the InAs deposition, we use the TMIn flux interruption while keeping the AsH$_3$ flux constant to tune the QDs formation. This step is further referred as "growth interruption". Then, the QD array was first covered with a $\SI{10}{\nano\meter}$-thick InP layer using TMIn and TBP sources. Then the temperature was raised to $\SI{610}{\degreeCelsius}$ for further growth of the InP layer. Then the layer of surface QDs was grown under conditions identical to the buried ones but without InP capping layers.

\subsection{Morphology and composition studies}

The morphology of buried In(As,P) layers and InAs/InP QDs was studied by means of high-angle annular dark-field scanning transmission electron microscopy (HAADF STEM) \cite{kadkhodazadeh2013} using an FEI TEM instrument equipped with a field emission electron gun and aberration correction on the probe forming lenses. 
The surface (non-buried) QDs density and QDs heights were investigated using the atomic force microscopy (AFM) Bruker Icon system operated in a tapping mode. Typical AFM images are given in Supplementary Information, section S1.

\subsection{ Optical studies}
For the optical experiments, the structures were kept in a helium-flow cryostat with temperature control in the range of $\SIrange{5}{300}{\kelvin}$ and pumped non-resonantly with a continuous-wave semiconductor laser diode line at $\lambda_{\mathrm{exc}}=\SI{640}{\nano\meter}$.
For high spectral and micrometer spatial-resolution photoluminescence ($\upmu$PL), the structures were optically excited through an infinity-corrected, high numerical aperture ($\mathrm{NA}=0.4$) microscope objective with $20\times$ magnification, focusing the laser spot to a diameter of $\SI{\sim2}{\micro\metre}$. For macro-PL we used a lens focusing the laser beam to a $\SI{\sim200}{\micro\metre}$ spot diameter.
The same objective or lens was used to collect the PL response and direct it for spectral analysis to a $\SI{0.3}{\meter}$-focal-length monochromator equipped with a liquid-nitrogen-cooled InGaAs multichannel array detector, and with a single-channel InAs detector measuring the macro-PL signal with the lock-in technique.

\section{ Results and discussion}

In the InAs/InP material system, there are two mechanisms to form epitaxial 2D In(As,P) layers. The first mechanism is the conventional epitaxial growth with both group III and V precursors supplied. The second one is a self-limited process of substitution of P atoms to As atoms during the annealing of InP in AsH$_3$ ambient \cite{gonzalez2002situ, yoon1999}. The V\textsuperscript{th} group atoms exchange is driven by the P desorption from the III-V surface at the temperatures above $\SI{350}{\degreeCelsius}$ \cite{gonzalez2002situ}. 
Here we employ both mechanisms and mainly discuss the MOVPE process, however, similar considerations would also be applicable in the case of molecular beam epitaxy \cite{gonzalez2002situ}.

\subsection{Growth modes of InAs on InP under As flux}

Under the standard growth conditions, when the InP surface is exposed to PH$_3$ ambient, the desorbing P atoms leave behind vacant adsorption sites, which are immediately filled with the phosphorus from the gas phase. In the case of annealing of the InP surface in the AsH$_3$ ambience, the As atoms bind to available indium atoms left vacant due to P desorption \cite{gonzalez2002situ}. The kinetics of As substitution of P atoms is temperature-dependent and the process is self-limited and reversible, as experimentally observed in Ref.~\cite{gonzalez2002situ}. 

\begin{figure}[H]
  \centering
  \includegraphics[width=\textwidth]{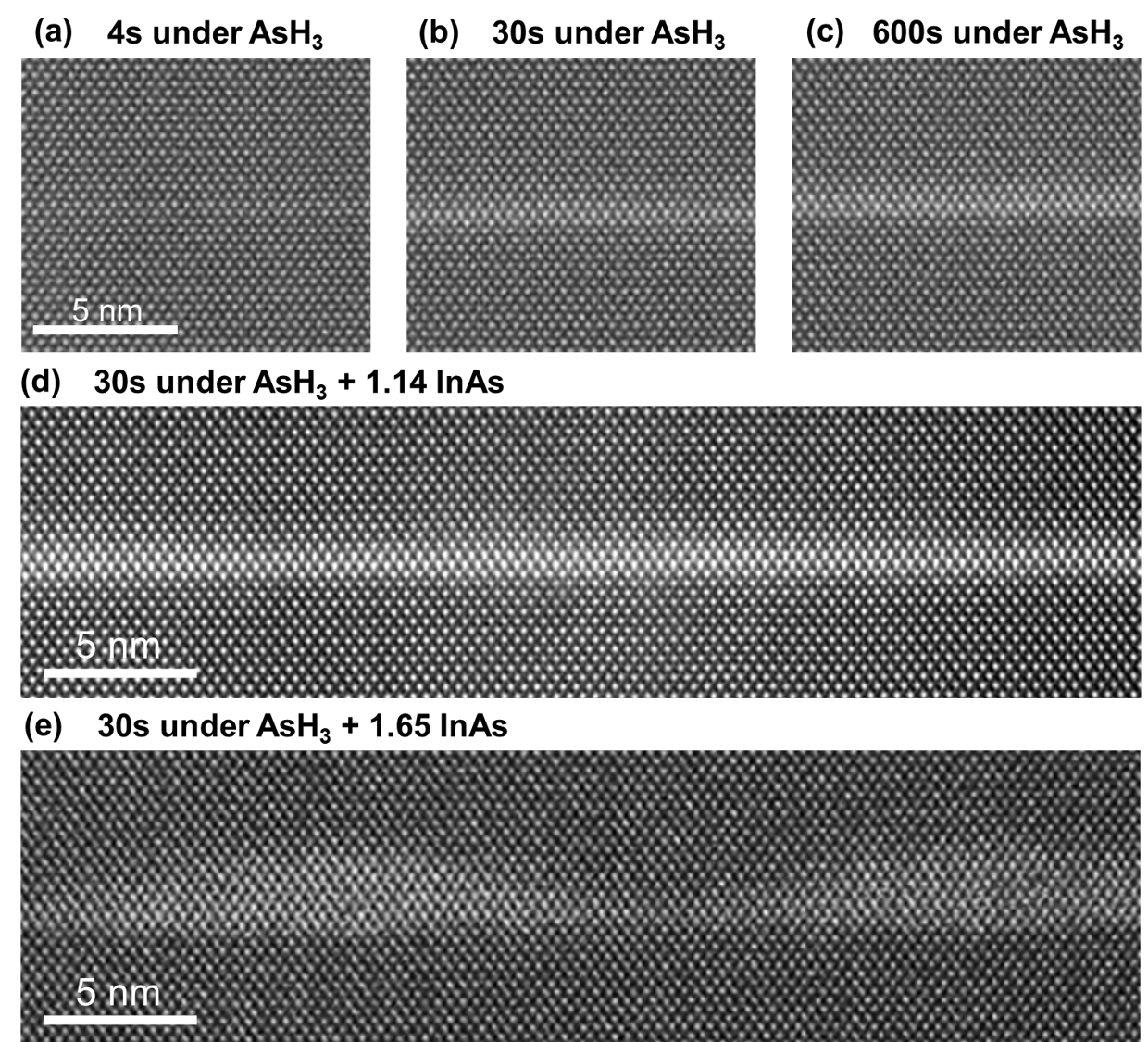} 
   \caption{ Cross-section HAADF STEM images of InAs layers formed after InP annealing in the AsH$_3$ ambient at $\SI{485}{\degreeCelsius}$ and AsH$_3$ flow of ${5.5 \cdot 10^{-5}}~\si{\mole\per\min}$ during: 
   (a) $\SI{4}{\second}$, (b)~$\SI{30}{\second}$, (c)~$\SI{600}{\second}$; and after $\SI{30}{\second}$ of annealing followed by additional deposition of (d) 1.14~ML and (e) 1.65~ML of InAs at V/III ratio of 5 and growth rate of 0.49~ML/s. }  
  \label{fig:WL}
\end{figure}

As mentioned earlier, from the model point of view, the transition between 2D and 3D growth becomes energetically favorable with the thickening of the WL. \subfigsrefL{WL}{a}{c} show the HAADF STEM images of the WLs formed in the three cases of InP substrate annealed at $\SI{485}{\degreeCelsius}$ under AsH$_3$ flow of ${5.5 \cdot 10^{-5}}~\si{\mole\per\min}$ for 4, 30 and 600 s, respectively. We have measured the thicknesses of the obtained In(As,P) layers based on the contrast in the HAADF STEM images and estimated the chemical composition of the layers from the displacement of crystal lattice planes along the [001] growth direction.

\subfigref{WL}{a} shows the STEM image of an approximately 2~ML-thick InAs$_{0.18}$P$_{0.82}$ layer formed after annealing for $\SI{4}{\second}$ subsequently covered by InP. Longer annealing promotes the formation of a thicker In(As,P) layer of higher As concentration.
\subfigsref{WL}{b}{c} show approx. 4~ML-thick layers of InAs$_{0.7}$P$_{0.3}$ and InAs$_{0.75}$P$_{0.25}$ formed after $\SI{30}{\second}$ and $\SI{600}{\second}$ of annealing in AsH$_3$, respectively. 
The In(As,P) layer thickness and As concentration after annealing for $\SI{600}{\second}$ is close to saturation under the given conditions. Although, no indication of plastic relaxation in the structures and QD formation could be found in the STEM images (\subfigsref{WL}{a}{c}). Therefore, the thickness and composition of the In(As,P) layer can be tuned by varying the duration of InP annealing under the AsH$_3$ flux. Meantime, under the considered experimental conditions, the thickness of the formed InAs(P) WL is not sufficient to initiate the formation of 3D islands. 

To approach the limit of 2D growth, the thickness and concentration of As in the In(As,P) layer need to be increased by further deposition of InAs. When the InAs layer exceeds the equilibrium thickness $h_{eq}$, the release of the accumulated elastic strain in the system leads to material redistribution, resulting in modulated thickness of the layer. These thickness modulations can be observed in the HAADF STEM image shown in \subfigref{WL}{d}.  
Namely, the deposition of $\sim$1.1~ML of InAs on top of the In(As,P) layer (formed after annealing in AsH$_3$ for 30~s, similar to \subfigref{WL}{b}), results in 1-2~ML thickness modulations on top of the $\sim$4~ML-thick In(As,P)with $\SI{\sim95}{\percent}$ of As (\subfigref{WL}{d}). As we discuss later in the AFM analysis, this thickness of InAs layer is enough to form very diluted arrays of QDs with surface densities below or far below $\SI{e8}{\per\square\centi\meter}$.  

Further increase of the InAs layer thickness beyond the critical value $h_c$
leads to formation of an array of coherently strained islands with their surface density exceeding $\SI{e10}{\centi\meter\tothe{-2}}$.
\subfigrefL{WL}{e} is an image of the self-assembled QDs after the deposition of 1.65~MLs of InAs in addition to $\SI{30}{\second}$ annealing under AsH$_3$. Ensembles of similar number density QDs can serve as the effective gain medium for lasing structures with a photonic crystal cavity \cite{yu2017, xue2016}. However, the fabrication of nonclassical photon sources based on a single QD requires much lower surface densities leading to a  substantial QD-to-QD separation \cite{Holewa2020PRB, holewa2022}. The partially consumed WL between the QDs in \subfigrefL{WL}{e} is about 2~ML thinner than the one after As-P exchange (\subfigref{WL}{b}). The WL composition in this sample is estimated to be InAs$_{0.9}$P$_{0.1}$ and the QD composition is close to pure InAs.

In summary, As-P exchange during the InP annealing under AsH$_3$ allows tuning the In(As,P) layer thickness and composition, although it is insufficient for InAs/InP QD formation under the considered growth conditions. Further deposition of InAs leads to QD formation. Deposition of less than 1.1~ML of InAs on top of $\sim$4~ML-thick InAs$_{0.7}$P$_{0.3}$ is required to exceed the equilibrium thickness, while the critical thickness in the considered structure is observed to be between 1.1 and 1.65~MLs deposited on top of In(As,P) layer.

Importantly, the formation of InAs/InP QDs strongly depends on growth parameters, which we discuss in detail in the following section. It is crucial to note here that TEM analysis generally is not capable of large area studies. Thus, our STEM results do not indicate the complete absence of QDs after additional deposition of 1.1~ML of InAs. It may just suggest that the QD density is much lower than that after adding 1.65~ML of InAs. These low-density QDs form at the WL thicknesses close to the critical value $h_c$.  We further explore this property by depositing of additional 1.1~ML (or less) InAs to obtain low-density near-critical QDs and tune their parameters.

\subsection{Influence of growth parameters on the properties of QDs}


Theoretical approaches\cite{li2010, dubrovskii2014, osipov2002stress} imply that the rate of QD nucleation is controlled by the superstress, which increases with the amount of deposited InAs. 
In the supercritical process (at  \(h \) above \(h_c\)), large superstress leads to fast nucleation and therefore rapid reduction of the WL thickness (and thus reduction of the superstress). This, in turn, limits the duration of the nucleation phase to a short time interval. In contrast, subcritical nucleation of QDs occurs at low superstress and continues throughout the growth.  
In this section, we investigate the formation of near-critical QDs (at $h \sim h_c$) and discuss ways to independently tune their size and surface density.

In the approximation of a short nucleation stage at WL thicknesses near the critical value, the surface density of nanoislands $N$ can be modeled as a step-like function \(\theta\) of the nominal thickness of the deposited InAs layer $h$  \cite{evans2006, dubrovskii2014, osipov2002stress}:

\begin{equation}
    N(h) = N_0  \theta (h)
    \label{eq_J}
\end{equation}

where $N_0$ is the saturated QD density for $h >> h_c$, which is dependent on the growth kinetics governed by the growth conditions.  Following the approach of Ref.~\cite{dubrovskii2014} we use the double-exponential expression for 
\( \theta(h) = 1 \exp(-\exp[\Gamma(h - h_c)/(h_c - h_eq) ] ) \), where \( \Gamma \) is twice the height of the nucleation barrier. Meantime, $N_0$ can be expressed as the function of the flux rate of the group III precursor $F_{In}$ and the rate of QD growth after nucleation $\nu$\cite{dubrovskii2014}:

\begin{equation}
   N_0 = n (F_{In}/ \nu)^{\alpha},
    \label{eq:eq_N0}
\end{equation}

with the crystal plane-specific coefficient $n$. The exponent $\alpha$ depends on the mechanism for the transport of the growth-limiting species.

For various InAs (001) surface reconstructions, the flux ratio is known to impact heavily the surface reconstruction and In adatom kinetics \cite{grosse2002, yeu2017}. At high V/III ratios, one can expect the QD formation on group V saturated surface with the growth rate limited by the surface diffusion of In adatoms. Several studies reported enhancement of In and Ga migration on the surfaces with excess of As \cite{shen1997, whitwick2009, shchukin2008thermodynamics}. In closer consideration, the diffusion of group III atoms on top of the WL is impacted by numerous surface processes and still needs to be fully understood. Therefore, to reflect the effect of In migration enhancement in our modeling, we assume that the rate of QD growth $\nu$ linearly increases with the As precursor flux. The simplified model approximation here is linear $\nu (F_{As})$ dependence applicable for the In-limited regime of QD growth.
In the case of the group III rich surface (low V/III ratio) the growth of QDs is limited by the transport of group V atoms from the gas phase and thus is assumed to be proportional to the As precursor flux. 
Next, we discuss the impact of the precursor flux ratio on the QD formation in further details.

\subsubsection{V/III flux ratio}

\begin{figure}[H]
  \centering
  \includegraphics[width=\textwidth]{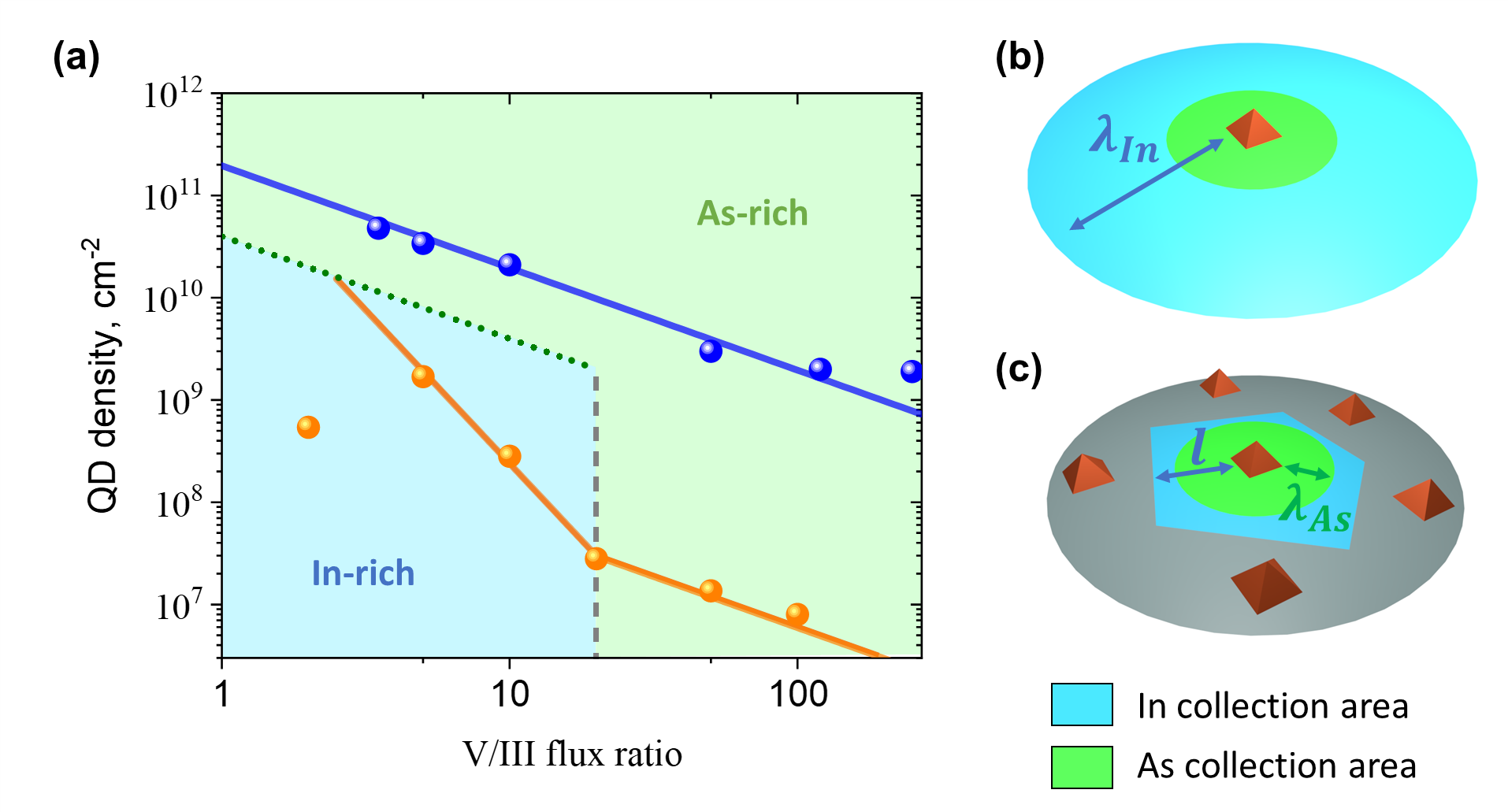} 
  \caption{ (a) Variation of QD surface density in supercritical (in blue) and near-critical (in orange) nucleation.  (b, c) Schematics for In and As collection areas in sparse and dense QD ensembles.}
  \label{fig:VIII}
\end{figure}

We observe that QD surface density has a different character of dependence on the flux ratio in dense and sparse ensembles of QDs, obtained after deposition of 1.2 and 0.8~MLs of InAs on top of the In(As,P) layer, correspondingly.
\subfigrefL{VIII}{a} shows more than two orders of magnitude change of the QD surface density ($\SIrange{e7}{e9}{\per\square\centi\meter}$)in sparse arrays of QDs while varying the As precursor flux to change the V/III ratio between 2 and 100 as illustrated in \subfigrefL{VIII}{a}.

As mentioned earlier, the exponent $\alpha$ is determined by the transport mechanism of the limiting growth species.  Under As-reach conditions the growth is limited by the transport of In to a QD, and in the excess of In – by the As species. In the case of In-limited (As-rich) growth \(\alpha = 1\), which corresponds to the surface diffusion of In atoms to the 3D nanoisland from a collection area much larger than the size of an island. This regime corresponds to the experimentally observed dependence at V/III ratios above 20 shown in \subfigrefL{VIII}{a}. In the case of the As-limited regime the measured dependence for low-density QDs corresponds to \(\alpha = 3\), which is consistent with the model regime of condensation from a 3D vapor in the ballistic regime \cite{dubrovskii2014}.

To delineate the In-limited and As-limited regimes, the transport of each growth species to a specific QD can be evaluated by multiplying the precursor flux density $F_i$ (with the index $i = In$ or $As$) by the area of the collection region around the QD $A_i$. Here $A_i = \pi \Lambda_i^2$ with  $\Lambda_i$ denoting the collection range for the corresponding type of the growth species. Thus, the equality \( F_{In} \Lambda_{In}^2=F_{As} \Lambda_{As}^2 \) formally defines the boundary between In-rich and As-rich conditions.

The collection range of In is limited either by the surface diffusion length \(\lambda_i\) or by the mean distance between QDs \( l\),  depending on which quantity is smaller. Thus, in terms of the notations we introduced above, \( \Lambda_i=\lambda_i \) in sparse ensembles and \(\Lambda_i = l \sim 1 / \sqrt{N}\) in dense arrays of QDs. Then, depending on the flux rates and QD surface density the following growth regimes can be distinguished:
\begin{enumerate}
    \item when \( N < \lambda_{In}^{-2} < \lambda_{As}^{-2} \) the fluxes of both types of growth species to a QD are given by the corresponding diffusion lengths \(\lambda_{As}\) and \(\lambda_{In}\) (as illustrated in \subfigrefL{VIII}{b}). Therefore, the flux ratio \(F_{As}/F_{In} = ( \Lambda_{In}/\Lambda_{As} )^2 \) delineates the In-rich (As-limited) regime from the As-rich (In-limited) one. This boundary is shown by the gray dashed line in \subfigrefL{VIII}{a}.
    \item when \( \lambda_{In}^{-2} < N <  \lambda_{As}^{-2} \) the As collection is limited by \( \lambda_{As} \) while the In collection - by \( l \) (see \subfigrefL{VIII}{c}). Thus, the boundary between two regimes is given by \( N =  \Lambda_{As}^{-2}   F_{In} / F_{As} \), shown by the green dotted line in \subfigrefL{VIII}{a}. Therefore, the As-limited growth is expected for low V/III ratios and QD surface densities within the blue-shaded region in \subfigrefL{VIII}{a}.
    \item when \(N > \lambda_{As}^{-2} > \lambda_{In}^{-2} \), both As and In collection lengths are limited by the distance between QDs and \( \Lambda_{In} = \Lambda_{As} = l \). Thus, the limiting type of growth species is the one with the minimal precursor flux ratio (in the assumption of complete pyrolysis). However, at very low V/III ratios segregation of In in droplets can be observed, therefore it is typically avoided for the SK QD growth. So, we consider here V/III ratios above 1 and therefore for high QD density the growth regime is always considered to be limited by In. 
\end{enumerate}

\begin{figure}[H]
  \centering
  \includegraphics[width=0.96\textwidth]{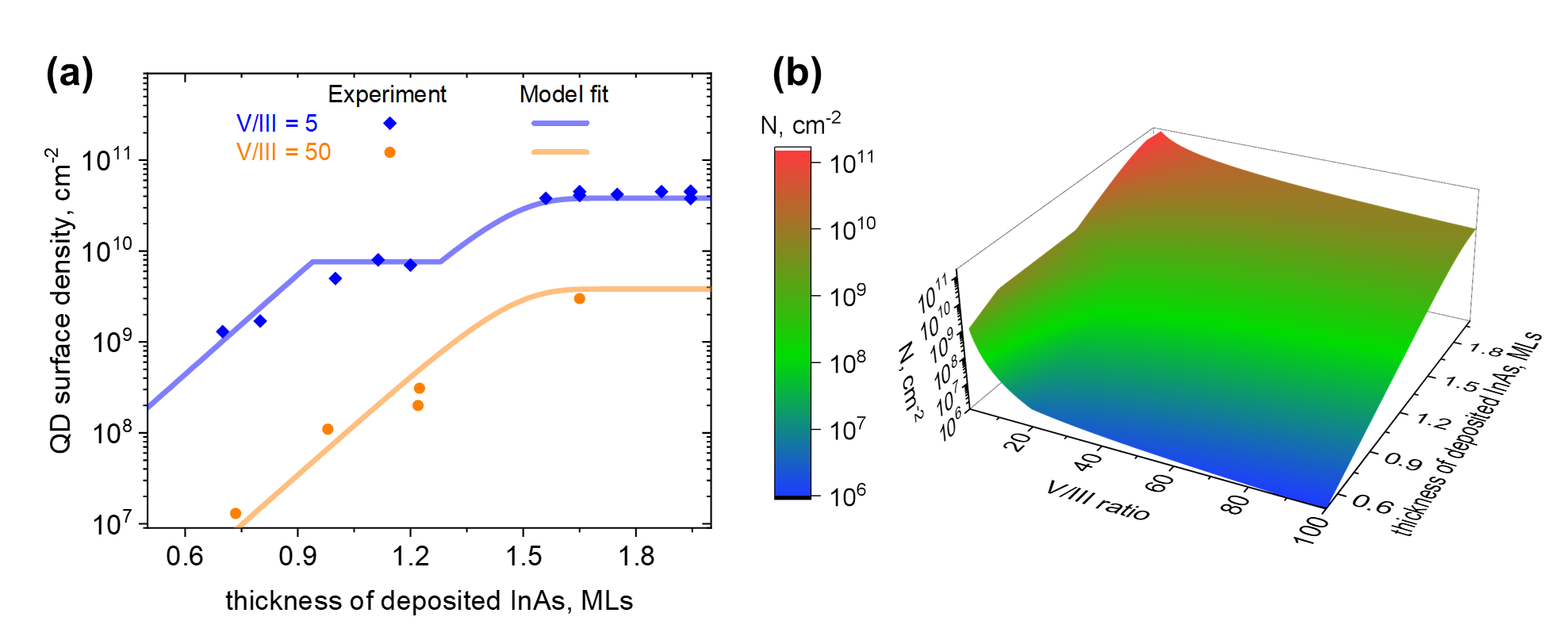} 
  \caption{ (a) Variation of QD surface density with the thickness of deposited InAs layer for V/III ratio of 5 and 50.  (b) Calculated QD surface density as the function of InAs layer thickness and V/III ratio.}
  \label{fig:VIII_2}
\end{figure}

\subfigrefL{VIII_2}{a} illustrates the variation of the QD surface density with the thickness of additionally deposited InAs layer in the cases of In-limited (V/III ratio of 50) growth and transition from As-limited to In-limited regimes (V/III ratio of 5). Solid lines show the model fit to experimental data. In the As-limited regime we assume that the nucleation of new QDs slows down considerably as soon as its surface density is large enough to limit the collection of In \cite{evans2006}. In this case we consider the growth of QDs to initiate in the As-limited regime and then proceed in the In-limited regime after the end of the nucleation phase. Therefore, we model the corresponding part of dependence with the constant QD surface density corresponding to the transition from As-limited to In-limited regimes. 

We use the model fits to extract the parameters in eq. (1-2) and generalize the experimental results to a wider range of V/III ratios and InAs layer thicknesses. \subfigrefL{VIII_2}{b} shows the color map for QD surface density calculated for V/III ratios from 1 to 100 and InAs layer thicknesses between 0.5 and 2 on top of the In(As,P) layer. One can note the sharp decrease in the QD density with the increase of V/III ratio around 20 which we associate with the As-limited nucleation regime which is specific for near-critical QDs formed at  \( h \) below 1 ML.

\subsubsection{Growth interruption}

\begin{figure}[H]
  \centering
  \includegraphics[width=\textwidth]{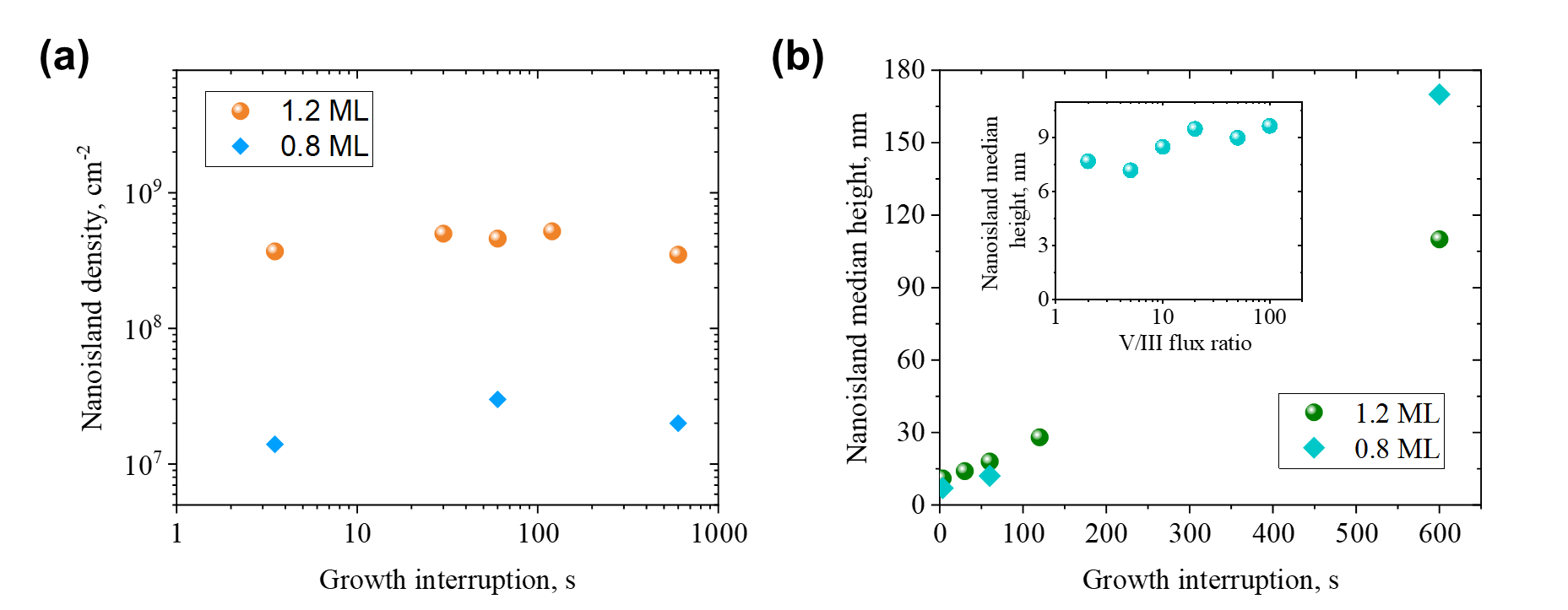} 
  \caption{ Surface density (a) and median height (b) of QDs as the functions of growth interruption duration after additional deposition of 0.8 and 1.2 ML of InAs. The inset in (b) shows the variation of QD density with V/III ratio. }  
  \label{fig:GI}
\end{figure}

For an arbitrary thickness of the InAs layer, we vary the duration of the growth interruption to further tune the properties of the QDs. \subfigrefL{GI}{a} shows the saturation of QD surface density as a function of the duration of the growth interruption.
After deposition of $1.2$~ML of InAs on top of the In(As,P) layer, the QD surface density remains around $(4 \pm 2) \cdot 10^8 \:  \mathrm{cm}^{-2}$ for all the growth interruptions up to $\SI{600}{\second}$. Deposition of slightly less of InAs  ($h=0.8$~MLs) results in order of magnitude lower QD densities saturated around $ (2 \pm 1) \cdot 10^7 \:  \mathrm{cm}^{-2}$.

While the surface density saturates, the height of the nanoislands increases with the duration of growth interruption as shown in \subfigrefL{GI}{b}.
We observe the increase of the median nanoisland height from 11 to 170 nm at $h=0.8$~ML and from 11 to 110 nm at $h = 1.2$~ML with the increase of growth interruption from $\SI{4}{\second}$ to $\SI{600}{\second}$.

Therefore, the size of low-density InAs islands can be controlled in a wide range without significant variation in surface density by tuning the duration of growth interruption. 
Importantly, the height of QDs changes insignificantly from $\SI{7}{\nano\meter}$ to $\SI{10}{\nano\meter}$, despite the two orders of magnitude variation of the QDs density with V/III ratio, as shown in the inset of \subfigrefL{GI}{b}.

\subsection{Optical properties of the QD system}

\begin{figure}[H]
  \centering
  \includegraphics[width=\textwidth]{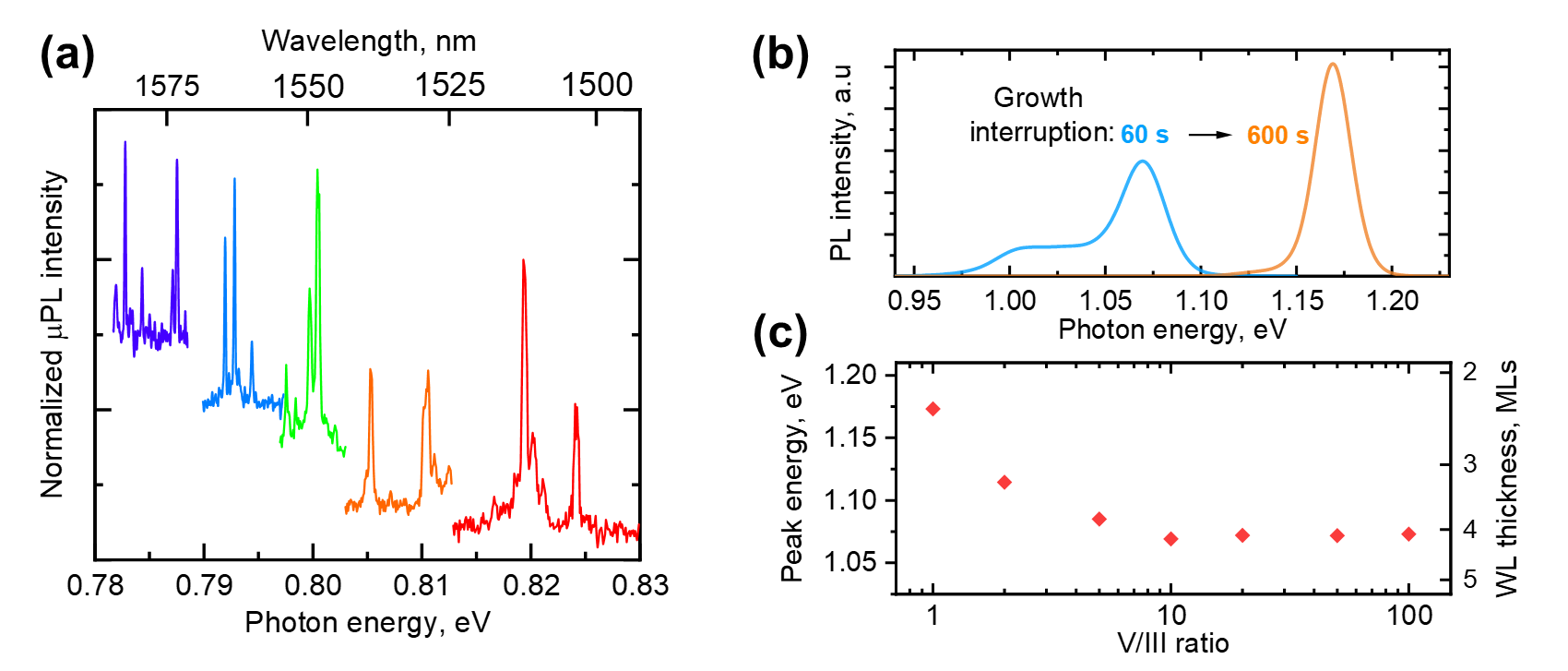} 
  \caption{ (a) Low temperature $\upmu$PL spectra of QDs acquired at $\SI{5}{\kelvin}$. (b) PL spectra of WL after 60 and $\SI{600}{\second}$ of growth interruption. (c) Measured energy of InAs$_x$P$_{1-x}$ WL peak as the function of V/III ratio for near-critical QDs. The right axis shows the corresponding WL thickness calculated for x = 0.75}  
  \label{fig:PL}
\end{figure}

In $\upmu$PL and macro-PL experiments we investigate the low-density QDs ($3 \cdot 10^8 \:  \mathrm{cm}^{-2}$) grown at V/III ratio of 50 and $h = 1.2$~ML.
In the $\upmu$PL studies of the samples with the growth interruption of $\SI{4}{\second}$, we observe the series of well-isolated lines corresponding to emission from single QDs in the spectral range of $\SIrange{1500}{1585}{\nano\meter}$, overlapping with the telecom C-band. \subfigrefL{PL}{a} shows the spectra of individual QDs acquired at $\SI{5}{\kelvin}$ and the optical pumping of $\SI{\sim1}{\micro\watt}$ (beam spot diameter of $\SI{\sim2}{\micro\meter}$).

In the macro-PL measurements performed in the broad spectral range with a $\SI{\sim200}{\micro\meter}$-large laser beam spot, only a PL signal from a WL is visible due to a significant disparity between 2D WL and 0D QDs density of states favorable to the former one. Therefore, PL emission is strongly dominated by carriers recombination in the WL. 

 WL peak energy in macro-PL measurements increases with the length of the growth interruption. \subfigrefL{PL}{b} shows a shift of the WL PL peak from $\SI{\sim1.07}{\electronvolt}$ to $\SI{\sim1.17}{\electronvolt}$ as the growth interruption changes from 60 to $\SI{600}{\second}$. The observed peak shift towards higher photon energies at longer growth interruptions is consistent with the model assumption of thinning of the WL during the growth of the nanoislands.

\subfigrefL{PL}{c} illustrates the variation of the WL PL peak energy measured for the series of samples grown at different V/III ratio. We observe the decreasing dependence at V/III ratios below 5 and saturation around $\SI{\sim1.07}{\electronvolt}$.

To quantify the observed changes of the PL signal, we calculate the energy of WL ground state transition (single-particle states) as a function of its thickness using the eight-band $\kp$ method in the \textit{nextnano} software~\cite{Birner2007}.
The WL is modeled as a homogeneous InAs$_x$P$_{1-x}$ quantum well at $\SI{5}{\kelvin}$ with $x = 0.75$ according to our TEM measurements.

The results of 8~$\kp$ calculations allow us to translate the measured values of peak energy to the estimated WL thickness (data on the right axis in \subfigrefL{PL}{c}).

As the WL peak position remains constant at V/III ratios above 5, it corresponds to saturation of the WL composition and thickness estimated to 4 MLs.
At V/III ratios below 5, the WL peak is shifted to higher energies corresponding to either lower WL thickness (down to 2 MLs) or lower As concentration.
This is consistent with the decrease of the QDs surface density (or even the absence of them) at V/III = 2 or lower.

\section{Conclusions}

We report a novel method of the formation of highly controllable low-density arrays of InAs/InP QDs via a near-critical nucleation process followed by In flux interruption under the growth temperature. With this approach, we have demonstrated how the surface density and mean size of QDs can be tuned almost independently from each other. Specifically the QD density can be tuned between $10^7$--$10^9$ $\si{\centi\meter\tothe{-2}}$ by choosing the V/III ratio between 2 and 100. 

\bibliographystyle{acm}
\bibliography{bibliography}

\end{document}